\documentstyle[emulateapj,psfig]{article}

\received{26 May 1998}
\accepted{23 June 1998}
\journalid{}{}
\articleid{}{}
\slugcomment{To appear in {\it the Astrophysical Journal, Letter}}

\lefthead{KOBAYASHI ET AL.}
\righthead{LOW-METALLICITY INHIBITION OF TYPE Ia SUPERNOVAE}

\def\gtsim {>\kern-1.2em\lower1.1ex\hbox{$\sim$}~}   
\def\ltsim {<\kern-1.2em\lower1.1ex\hbox{$\sim$}~}   

\begin{document}

\title{LOW-METALLICITY INHIBITION OF TYPE Ia SUPERNOVAE \\
AND \\ GALACTIC AND COSMIC CHEMICAL EVOLUTION}
\author{Chiaki KOBAYASHI$^1$, Takuji TSUJIMOTO$^2$, Ken'ich NOMOTO$^{1}$, 
Izumi HACHISU$^3$, Mariko KATO$^4$}
\affil{$^1$ Department of Astronomy \& Research Center for the Early Universe, 
School of Science,
University of Tokyo,\\ Bunkyo-ku, Tokyo 113-0033, Japan;
chiaki@astron.s.u-tokyo.ac.jp, nomoto@astron.s.u-tokyo.ac.jp}
\affil{$^2$ National Astronomical Observatory, Mitaka, Tokyo 181-8588, Japan;
tsuji@misty.mtk.nao.ac.jp}
\affil{$^3$ Department of Earth Science and Astronomy, 
College of Arts and Sciences, University of Tokyo, \\
Meguro-ku, Tokyo 153-8902, Japan;
hachisu@chianti.c.u-tokyo.ac.jp}
\affil{$^4$ Department of Astronomy, Keio University, 
Kouhoku-ku, Yokohama 223-8521, Japan;
mariko@educ.cc.keio.ac.jp}

\begin{abstract}
We introduce a metallicity dependence of Type Ia supernova (SN Ia) rate
into the Galactic and cosmic chemical evolution models. 
In our SN Ia progenitor scenario, the accreting white dwarf (WD) blows a
strong wind to reach the Chandrasekhar mass limit.
If the iron abundance of the progenitors is as low as [Fe/H] $\ltsim -1$,
then the wind is too weak for SNe Ia to occur.
Our model successfully reproduces the
observed chemical evolution in the solar neighborhood.
We make the following predictions which can test this metallicity effect:
1) SNe Ia are not found in the low-iron abundance environments
such as dwarf galaxies and the outskirts of spirals.
2) The cosmic SN Ia rate drops at $z \sim 1-2$ due to the low-iron abundance,
which can be observed with the Next Generation Space Telescope.
At $z \gtsim 1-2$, SNe Ia can be found only in the environments where
the timescale of metal enrichment is sufficiently short 
as in starburst galaxies and ellipticals.

The low-metallicity inhibition of SNe Ia 
can shed new light on the following issues: 
1) The limited metallicity range of the SN Ia progenitors would imply that
``evolution effects'' are relatively small for the use of high redshift SNe Ia
to determine the cosmological parameters.
2) WDs of halo populations are poor producers of SNe Ia,
so that the WD contribution to the halo mass is not constrained from
the iron abundance in the halo.
3) The abundance patterns of globular clusters and field stars 
in the Galactic halo lack of SN Ia signatures in spite of
their age difference of several Gyrs,
which can be explained by the low-metallicity inhibition of SNe Ia.
4) It could also explain why the SN Ia contamination is not seen
in the damped Ly$\alpha$ systems for over a wide range of redshift.
\end{abstract}

\keywords{Abundances --- binaries: close --- Cosmology: general --- 
Galaxy: evolution --- galaxies: evolution --- stars: supernovae}

\section{INTRODUCTION}

There exist two distinct types of supernova explosions: One is 
Type II supernovae (SNe II), which are the core collapse-induced explosions
of short-lived massive stars ($\gtsim \, 8M_\odot$)
and produce more O and Mg relative to Fe (i.e., [O/Fe] $>0$), and the other is
Type Ia supernovae (SNe Ia), which are the thermonuclear explosions 
of accreting white dwarfs (WDs) in close binaries 
and produce mostly Fe and little O.
The exact companion stars of the WDs have not been identified 
but must be relatively long-lived low mass stars.

The role of these two types of supernovae in the chemical evolution of
galaxies can be seen in the abundance ratios of stars with different
metallicities, most notably in the [O/Fe]-[Fe/H] relation.  
Metal poor stars with [Fe/H] $\ltsim -1$ have [O/Fe] $\sim 0.45$ on the
average (\cite{nis94}; \cite{gra91}),
while disk stars with [Fe/H] $\gtsim -1$ show a decrease in [O/Fe] 
with increasing metallicity (\cite{edv93}; \cite{bar89}; \cite{gra91}).  
Such an evolutionary change in [O/Fe] against [Fe/H] has been explained with 
the early heavy element production by SNe II and
the delayed enrichment of Fe by SNe Ia (\cite{mat86}).  
Conversely, chemical evolution models can constrain the nature of
still uncertain progenitor systems of SNe Ia;  
for example, Yoshii, Tsujimoto, \& Nomoto (1996) 
estimated the lifetime of SN Ia progenitors as long as $0.5-3$ Gyr.

SNe Ia have been discovered up to $z\sim0.93$ by 
Supernova Cosmology Project (\cite{per97}) and High-z Supernova Search
Team (\cite{gar98}).
They have given the SN Ia rate at $z\sim0.4$ (\cite{pai96}) but will
provide the SN Ia rate history over $0<z<1$.  With the Next Generation
Space Telescope, both SNe Ia and II will be observed through $z\sim 4$.
In theoretical approach, the cosmic SN Ia rate as a function of redshift
has been constructed for a cosmic star formation rate (SFR)
(\cite{rui98}; \cite{yun98}; \cite{sad98}; \cite{mad98}).
The comparison between the model prediction and observations can constrain
the lifetime of the progenitor systems of SNe Ia.

The progenitors of the majority of SNe Ia are most likely 
the Chandrasekhar (Ch) mass WDs (e.g., \cite{nom97a} for a recent review),
although the sub-Ch mass models might correspond to some peculiar 
sub-luminous SNe Ia.
The early time spectra of the majority of SNe Ia are
in excellent agreement with the synthetic spectra of the Ch mass models,
while the spectra of the sub-Ch mass models are too blue to
be compatible with observations (\cite{hof96}; \cite{nug97}).
For the evolution of accreting WDs toward the Ch mass,
two scenarios have been proposed: 
One is a double degenerate (DD) scenario, i.e., merging of double C+O WDs
with a combined mass surpassing the Ch mass limit 
(\cite{ibe84}; \cite{web84}), and the other is  
a single degenerate (SD) scenario,
i.e., accretion of hydrogen-rich matter via mass transfer from
a binary companion (e.g., \cite{nom94} for a review).  The issue
of DD vs. SD is still debated (e.g., \cite{bra95} for a review),
although theoretical modeling has indicated that 
the merging of WDs does not make typical SNe Ia (\cite{sai85}, 1998).

For the SD scenario, a new evolutionary model has been proposed by
Hachisu, Kato, \& Nomoto (1996, 1998; hereafter HKN96 and HKN98, respectively)
and Hachisu \& Kato (1998; hereafter HK98).
HKN96 have shown that if the accretion rate exceeds a certain limit,
 the WD blows a strong wind and burns hydrogen steadily to increase 
the WD mass (see section 2).
HKN98 have further invoked the effect of stripping-off of the envelope from
the companion by the strong wind, and shown that the WD can reach the Ch mass 
for much wider binary parameter space than found by HKN96,
 Li \& van den Heuvel (1997), and Yungelson \& Livio (1998);
the allowed parameter space may be large enough to account for 
the SN Ia frequency.
Moreover, HK98 have found an important metallicity effect;
if the iron abundance of the accreted matter is as low as
[Fe/H] $\ltsim -1$, the WD wind is too weak to increase the WD mass
through the Ch mass.

In the present {\sl Letter}, 
we apply the above two scenarios to the chemical evolution models,
and compare the cases with and without the metallicity effect on SNe Ia.
We have found that the model for SD scenario with the metallicity effect
is significantly better to reproduce the evolutionary change in [O/Fe] and
other properties (Section 3).  
Using the metallicity dependent SN Ia rate, we
make a prediction for the cosmic supernova rate history (Section 4).
In section 5 we discuss other implications on the Galactic halo objects, 
damped Ly$\alpha$ (DLA) systems and cosmology.

\section{TYPE Ia SUPERNOVA PROGENITOR SYSTEM}

     Our SD scenario has two progenitor systems: One is a red-giant
(RG) companion with the initial mass of $M_{\rm RG,0} \sim 1 M_\odot$
and the orbital period of tens to hundreds days (HKN96; HKN98). The
other is a near main-sequence (MS) companion with the initial mass of
$M_{\rm MS,0} \sim 2-3 M_\odot$ and the period of several tenths of a
day to several days (\cite{li97}; HKN98).  In these SD scenarios,
optically thick winds from the mass accreting WD play an
essential role in stabilizing the mass transfer and escaping from
forming a common envelope.  The optically thick winds are driven by a
strong peak of OPAL opacity at $\log T ({\rm K}) \sim 5.2$ 
(e.g., \cite{igl93}).  Since the peak is due to iron lines, the optically
thick winds depend strongly on the metallicity (HK98).

Figure 1 shows the metallicity dependence of the
optically thick winds.  
The strong winds are possible only for the
region above the dashed line.  The term ``weak'' implies that the
wind velocity at the photosphere does not exceed the escape velocity
there, that is, it cannot blow the accreted matter off the WD. 
For the metallicity as small as $Z=0.001$, 
the opacity peak at $\log T \sim 5.2$ is very weak, being
smaller than the peak of helium lines at $\log T \sim 4.6$.
Then, the wind is driven by the
helium line peak rather than the iron line peak, which we call
``He wind'' instead of ``Fe wind''.  Since only the initial WD
 mass of $M_{\rm WD,0} \leq 1.2 M_\odot$ can produce an SN Ia (\cite{nom91}), 
SN Ia events occur only for the progenitors with [Fe/H] $\gtsim -1.1$,
which is adopted in our chemical evolution model.

Figure 2 shows the SN Ia regions in the diagram of the initial
orbital period vs. the initial mass of the companion star for the
initial WD mass of $M_{\rm WD,0}=1.0 M_\odot$ (see HK98 for other
$M_{\rm WD,0}$).  In these regions, the accretion from the companion
star increases the WD mass successfully through the occurrence of SN Ia.
The dashed line shows the case of solar abundance ($Z=0.02$),
while the solid line shows the much lower metallicity case of $Z=0.004$.
The size of these regions clearly demonstrate the metallicity effect, 
i.e., SN Ia regions are much smaller for smaller metallicity. 
The initial mass ranges of the companion stars for $Z=0.004$ are 
$0.9 M_\odot \ltsim M_{\rm RG,0} \ltsim 1.5 M_\odot$ for the WD+RG system and
$1.8 M_\odot \ltsim M_{\rm MS,0} \ltsim 2.6 M_\odot$ for the WD+MS system.

\section{THE CHEMICAL EVOLUTION IN THE SOLAR NEIGHBORHOOD}

We use the chemical evolution model which allows the material infall from
outside the disk region.  For the infall rate, we adopt a formula
which is proportional to $t\exp[-t/\tau]$ with a infall timescale of
$\tau=5$ Gyr (Yoshii et al. 1996). 
The Galactic age is assumed to be 15 Gyr.
 The SFR is assumed to be
proportional to the gas fraction with a constant rate coefficient of
$0.37$ Gyr$^{-1}$.  For the initial mass function (IMF), we assume a
power-law mass spectrum with a Salpeter slope of $x=1.35$ in the range of 
$0.05M_\odot \leq M \leq 50M_\odot$ (\cite{tsu97}).
We take the nucleosynthesis yields of SNe Ia and II from 
Tsujimoto et al. (1995; see Nomoto et al. 1997b,c for details)
and the metallicity dependent main-sequence lifetime from Kodama (1997).

For the SD scenario, the lifetime of SNe Ia is determined from the
main-sequence lifetime of the companion star. 
We adopt the initial mass ranges of the binary companion stars 
obtained in section 2.
The distribution function of the
initial mass of the companions is taken from the mass ratio
distribution in binaries (\cite{dug91}), which is approximated by a
power-law mass spectrum with a slope $x=0.35$.  The fraction of
primary stars of $3-8 M_\odot$ which eventually produce SNe Ia is
set to be $0.04$ for both the WD+MS and the WD+RG systems,
adjusted to reproduce the chemical evolution in the solar neighborhood.
For the DD scenario 
we adopt the distribution function of the lifetime of SNe Ia by
Tutukov \& Yungelson (1994), majority of which is $\sim 0.1-0.3$ Gyr.

Figure 3 shows the evolutionary change in [O/Fe] for three SN Ia
models. The dotted line is for the DD scenario.  
The other lines are for our SD scenario with (solid line) and without
(dashed line) the metallicity effect on SNe Ia. 
The results are as follows.

\begin{itemize}
\item In the DD scenario the lifetime of the majority of SNe Ia is 
shorter than $0.3$ Gyr. 
Then the decrease in [O/Fe] starts at [Fe/H] $\sim -2$, which is too early
compared with the observed decrease in [O/Fe] starting at [Fe/H] $\sim -1$.
\item For the SD scenario with no metallicity effect, the companion
star with $M \sim 2.6 M_\odot$ evolves off the main-sequence to give
rise to SNe Ia at the age of $\sim 0.6$ Gyr. The resultant decrease in
[O/Fe] starts too early to be compatible with the observations. 
\item For the metallicity dependent SD scenario, 
SNe Ia occur at [Fe/H] $\gtsim -1$,
which naturally reproduce the observed break in [O/Fe] at [Fe/H] $\sim -1$.
\end{itemize}

We also perform the Kolmogorov-Smirnov (KS) test to estimate how well
the predicted abundance patterns agree with the observed one
(\cite{edv93}), and tabulate the resultant probabilities in Table 1.
The KS test is applied to the abundance distribution function in
the range of $-1.15 \leq \mbox{[Fe/H]} \leq 0.45$ and to the
evolutionary behavior of [O/Fe] in the range of $-0.85 \leq
\mbox{[Fe/H]} \leq 0.02$.  The predicted abundance distribution
functions are identical with the observational data with more than
$84\%$ probability for both the DD and SD scenarios.  For our SN Ia model
with the metallicity effect, the calculated [O/Fe] against [Fe/H] fits well
to the observations with $85\%$ probability.  But when we
neglect the metallicity effect, the probability decreases to $57\%$.
For the DD scenario the calculated [O/Fe] is rejected with $82\%$ probability.

\section{COSMIC SUPERNOVA RATE}

We apply our SN Ia progenitor scenario to predict the cosmic supernova
rate history and the cosmic chemical evolution corresponding to the
observed cosmic SFR (\cite{mad96}; \cite{con97}).
The photometric evolution is calculated with the spectral synthesis
population database taken from Kodama (1997).  
We adopt $H_0=50$ km s$^{-1}$ Mpc$^{-1}$, $\Omega_0=0.2$, $\lambda_0=0$,
and the redshift at the formation epoch of galaxies $z_{\rm f}=5$.
We use the initial comoving density of gas
 $\Omega_{{\rm g}\infty}=2\times10^{-3}$ (\cite{pei95}).

Figure 4 shows the cosmic supernova rate per $10^{10} L_\odot$ per
century (SNu).  The long dashed line is for SNe II and the other lines for 
SNe Ia with (solid line) and without (dashed line) the metallicity effect.  
If we do not include the metallicity effect, the SN Ia rate is
almost flat from the present to higher redshift, 
and decreases toward the formation epoch of galaxies.
If we include the metallicity effect, the SN Ia rate drops at 
$z \sim 1.2$, where the iron abundance of the gas in the universe
is too low (i.e., [Fe/H] $\ltsim -1$) 
for the progenitors of SNe Ia to make explosions. 
The redshift where the SN Ia rate drops
is determined by the speed of the chemical enrichment, which depends on 
the effect of dust extinction on the cosmic SFR (\cite{pet97}),
cosmology, galaxy formation epoch and the initial gas density.
Taking into account these uncertainties, the break 
in the SN Ia rate occurs at $z=1-2$.
We should note that galaxies being responsible for the global SFR have
different heavy-element enrichment timescale,
some of which are starburst galaxies and ellipticals 
achieving [Fe/H] $\sim -1$ in much shorter timescale than in Figure 4.
In such galaxies SNe Ia can occur even at $z \gtsim 1-2$.

\section{CONCLUSIONS AND DISCUSSION}

We introduce a metallicity dependence of the SN Ia rate
in the Galactic and cosmic chemical evolution models. 
In our scenario involving a strong wind from WDs, 
little SNe Ia occur at [Fe/H] $\ltsim -1$. 
Our model successfully reproduces the observed chemical evolution
in the solar neighborhood.
We make the following predictions which can test this metallicity effect.
1) SNe Ia are not found in the low-iron abundance environments
such as dwarf galaxies and the outskirts of spirals.
2) The cosmic SN Ia rate drops at $z \sim 1-2$ due to the low-iron abundance,
which can be observed with the Next Generation Space Telescope.
At $z \gtsim 1-2$, SNe Ia can be found only in the environments where
the timescale of metal enrichment is sufficiently short 
as in starburst galaxies and ellipticals.

The low-metallicity inhibition of SNe Ia
can shed new light on the following issues: 

1) It imposes a limit 
on the metallicity range of the SN Ia progenitors.
This would imply that ``evolution effects'' are relatively small 
for the use of high redshift SNe Ia
to determine the cosmological parameters.

2) Microlensing experiments (\cite{alc97}) suggest
the WD-dominated Galactic halo.
The existence of so many WDs 
results in too much iron enrichment from SNe Ia (\cite{can97}).
However, WDs of halo populations are poor producers of SNe Ia, so that
the iron abundance of the halo may not put the strong constraint 
on the contribution of WDs.
Note, however, that there exist several other arguments 
against the WD-dominated halo (e.g., \cite{cha95}; \cite{gib97}).

3) The Galactic halo is a low-metallicity system with [Fe/H] $\ltsim -1$
and have an abundance pattern of genuine SN II origin, i.e.,
the overabundances of $\alpha$-elements relative to Fe as [$\alpha$/Fe] $>0$
(e.g., \cite{wee89}).
However there exist age differences of several Gyrs among the clusters
(\cite{cha96}) as well as field stars (\cite{sch89}).
Since the shortest lifetime of SNe Ia is $\sim 0.6$ Gyr 
for the MS+WD close binary systems,
SN Ia contamination would be seen in [$\alpha$/Fe] 
if there were no metallicity effect on SNe Ia.
This apparent discrepancy 
between the age difference and the high [$\alpha$/Fe] 
can be resolved by the low-metallicity inhibition of SNe Ia.

4) Similar interpretation holds also for DLA systems.
The DLA systems observed at $0.7<z<4.4$ have [Fe/H]$=-2.5$ to $-1$ 
and indicate [$\alpha$/Fe] $>0$ (\cite{lu96}).
Lu et al. (1996) suggested there may exist the
age-metallicity relation in DLA systems, which implies that DLA
systems grown through a common chemical history spanning over several Gyrs. 
If so, the SN II-like abundance pattern in DLA systems needs 
the introduction of the metallicity dependent SN Ia rate 
to avoid the contamination of SN Ia products.

\acknowledgments

This work has been supported in part by the grant-in-Aid for
Scientific Research (05242102, 06233101, 08640336, 08640321, 09640325)
and COE research (07CE2002) of the Ministry of Education, Science,
Culture, and Sports in Japan.
We would like to thank Tadayuki Kodama for providing us with 
the database of simple stellar population spectra,
and David Branch for many useful comments to improve the paper.


\begin{deluxetable}{lcc}
\footnotesize
\tablewidth{0pt}
\tablecaption{Probabilities given by the Kolmogorov-Smirnov test}
\tablehead{
\multicolumn{1}{c}{SN Ia progenitor model} &
\multicolumn{2}{c}{Probability} \\
\cline{2-3}
 & \colhead{abundance ratio} & \colhead{abundance distribution}
}
\startdata
SD with the metallicity effect    & $85.0 \,\%$ & $89.5 \,\%$ \nl
SD without the metallicity effect & $57.2 \,\%$ & $84.8 \,\%$ \nl
DD scenario                       & $17.6 \,\%$ & $86.9 \,\%$ \nl
\enddata
\end{deluxetable}

\newpage

\begin{figure}[ht]
\centerline{\psfig{figure=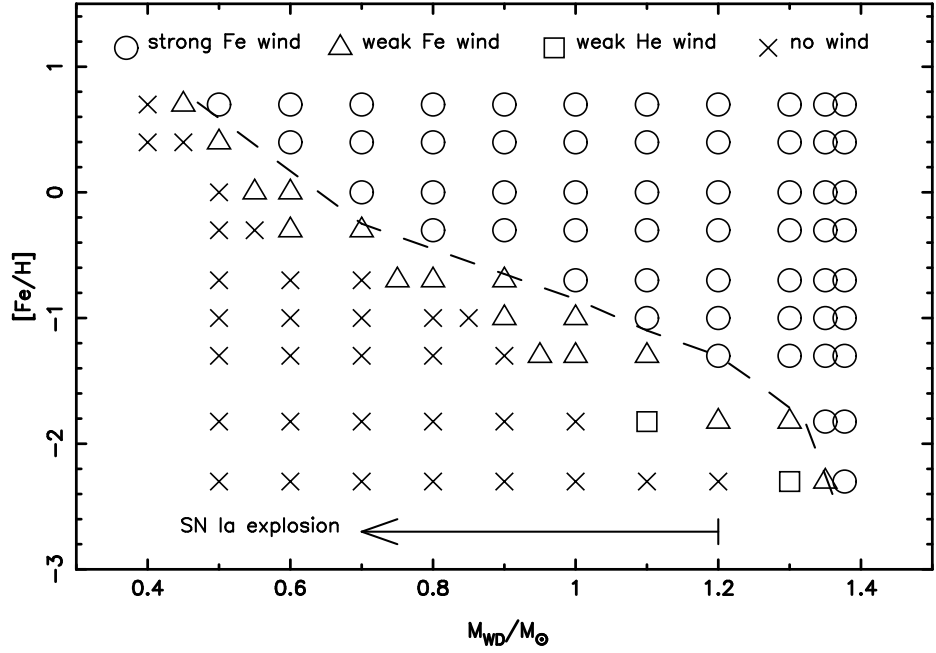,width=13.3cm}}
\caption[fig1.ps]{
Metallicity dependence of optically thick winds is shown 
in WD mass vs. metallicity diagram.  We regard the wind 
as ``strong'' if the wind velocity at the photosphere exceeds 
the escape velocity but ``weak'' if the wind velocity is lower than 
the escape velocity.  The term of ``He'' or ``Fe'' wind denotes 
that the wind is accelerated by the peak of iron lines 
near $\log T ({\rm K}) \sim 5.2$ or of helium lines 
near $\log T ({\rm K}) \sim 4.6$.  The dashed line indicates 
the demarcation between the ``strong'' wind and the ``weak'' wind.
}
\end{figure}

\begin{figure}[ht]
\centerline{\psfig{figure=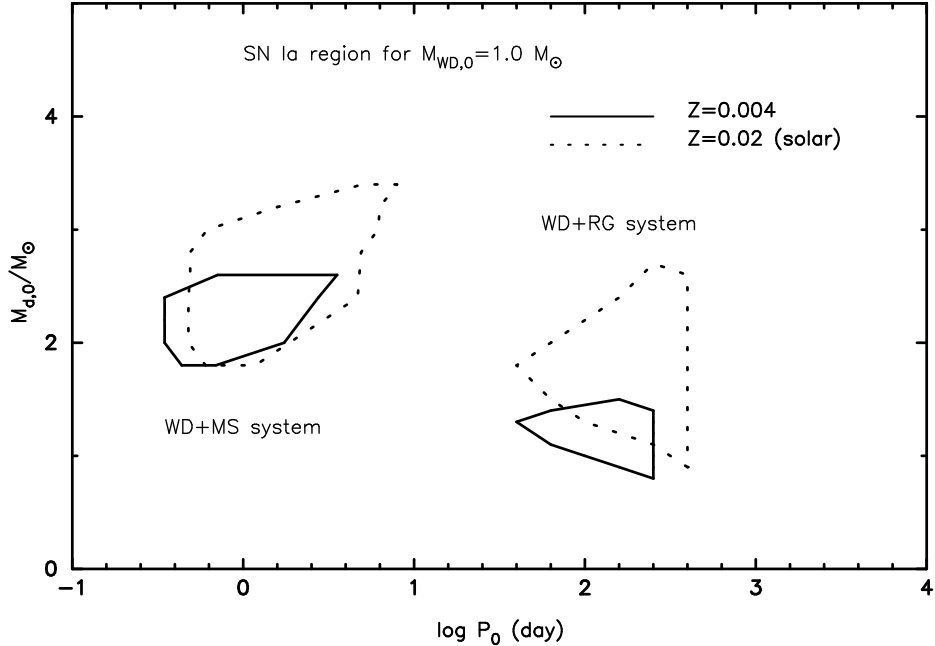,width=13.3cm}}
\caption[fig2.ps]{
The regions of SNe Ia is plotted in 
the initial orbital period vs. the initial companion 
mass diagram for the initial WD mass of $M_{\rm WD,0}=1.0 M_\odot$.  
The dashed and solid lines represent the cases of 
solar abundance ($Z=0.02$) and much lower metallicity of $Z=0.004$,
 respectively. 
The left and the right regions correspond to 
the WD+MS and the WD+RG systems, respectively. 
}
\end{figure}

\newpage
\begin{figure}[ht]
\centerline{\psfig{figure=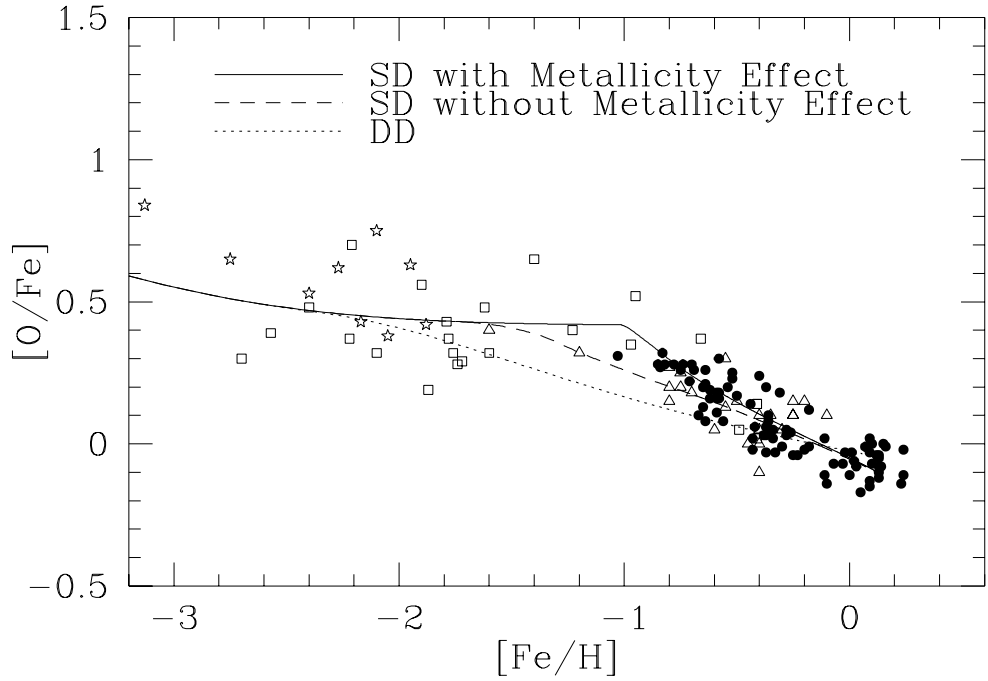,width=13.3cm}}
\caption[fig3.ps]{
The evolutionary change in [O/Fe] against [Fe/H] for three SN Ia models. 
The dotted line is for the DD scenario 
where SNe Ia occur by merging of two WDs at a rate given 
by Tutukov \& Yungelson (1994). 
The other lines are for our SD scenario with (solid line) and 
without (dashed line) the metallicity effect on SNe Ia.
Observational data sources: 
filled circles, Edvardsson et al. (1993);
open triangles, Barbuy \& Erdelyi-Mendes (1989);
stars, Nissen et al. (1994);
open squares, Gratton (1991).
}
\end{figure}

\begin{figure}[ht]
\centerline{\psfig{figure=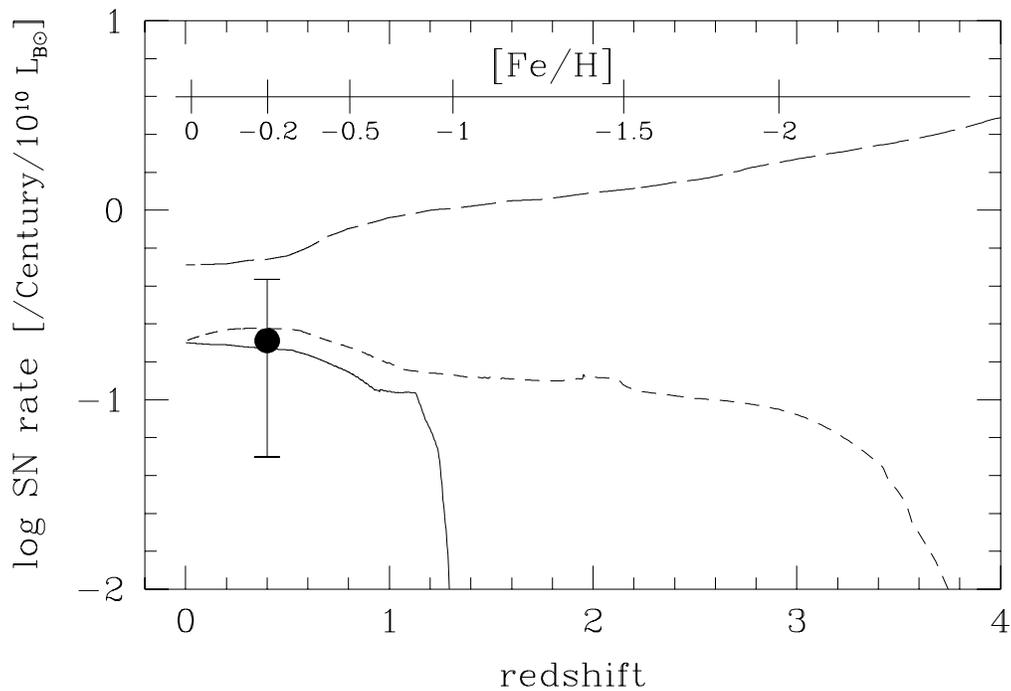,width=13.3cm}}
\caption[fig4.ps]{
The cosmic supernova rate per $10^{10} L_\odot$ per $\mbox{century}$ (SNu).
The long-dashed line is for SNe II and the other lines for SNe Ia 
with (solid line) and without (dashed line) the metallicity effect.
The filled circle is the observed SN Ia rate at $z \sim 0.4$ 
(Pain et al. 1996).
The iron-abundance scale in the abscissa is 
calculated with the metallicity effect on SNe Ia.
}
\end{figure}

\end{document}